# The global critical current effect of superconductivity

Heng Wu[1,2*], Yaojia Wang[1,2], Mazhar N. Ali[1,2]

[1]Department of Quantum Nanoscience, Faculty of Applied Sciences, Delft University of Technology, Lorentzweg 1, 2628 CJ, Delft, The Netherlands

[2]Kavli Institute of Nanoscience, Delft University of Technology, Lorentzweg 1, 2628 CJ, Delft, The Netherlands

*Corresponding author. Email: wuhenggcc@gmail.com

**Abstract**

Superconductivity has been investigated for over a century, but there are still open questions about what determines the critical current[1]; the maximum current a superconductor can carry before switching to its normal state. For a given superconductor, the zero-field critical current is widely believed to be determined by its inherent properties and be related to its critical magnetic field[2,3]. Here, by studying superconducting polycrystalline films, single crystal flakes, and layered heterostructures, we find that the critical current of a superconductor can vary with measurement configuration. It can be influenced, or even determined, by adjacent superconducting segments along the applied current trajectory, whereas the critical magnetic field and critical temperature remain unaffected. This "global critical current effect" both reveals the need to revisit fundamental theory describing superconductivity, as well as implies that superconductors can transfer critical current related properties to each other. We demonstrate this by designing and fabricating a simple superconducting structure that transferred a superconducting diode effect from one segment to another segment which could not manifest the effect on its own. This observation merits a reconsideration of contemporary superconducting circuit design, and developing a full understanding will lead to a new paradigm of superconducting electronics.

**Introduction**

The critical current ($I_c$), one of the basic traits of a superconductor, is an essential consideration for use of superconductors in many applications[4], such as power transmission lines, electromagnets, as well as superconducting electronics and circuits including devices based on the Josephson effect[5], like quantum computing and sensing[6]. Recently, critical current related phenomena such as superconducting diode effect[7–9], where the $I_c$ in one direction is not equal to the $I_c$ in the opposite direction, has attracted large attention and has caused the community to reevaluate the understanding of critical currents in the context of broken symmetries and nonlinear responses[10–13]. Although the critical current of a superconductor was noticed immediately upon the discovery of superconductivity by Onnes[1], new discoveries are still being made and the fundamental mechanism of how current breaks the superconductivity is not yet fully understood[1,14].

Throughout the history, many efforts have been made to understand the origin of the critical current in a superconductor. Due to the fact that current can induce a magnetic field, early understanding of the critical current without an external magnetic field (aka zero field) was described by Silsbee's hypothesis[15], where in type-I superconductors, the critical current is simply the current that induces a magnetic field (aka self field) large enough to reach the critical magnetic field ($B_c$). In this case, the critical current density ($J_c$) of a superconductor takes the form of $J_c = \frac{B_c}{\mu_0 \lambda}$, where $\mu_0$ is the vacuum permeability, $\lambda$ is the London penetration depth[1–3]. Later, in Ginzburg and Landau's phenomenological theory, $B_c^2/2\mu_0$ is used as the difference in the Gibbs free energy density between the normal and superconducting states, again tying the $J_c$ with the $B_c$[2,3]. In type II superconductors without external magnetic field, flux vortices can only be nucleated through the self field of the applied current, and $J_c$ is found to be proportional to the lower critical magnetic field $B_{c1}$[1,14,17]. In the presence of an external magnetic field, it has been shown that $J_c$ is dependent on the pinning force of the vortices, which is dependent on the geometry and microstructure of the superconductor[18,19].

Therefore, it is generally believed that the zero-field critical current density of a superconductor is associated with its critical magnetic field, and that for a superconductor with a given geometry and internal microstructure, its $J_c$ is fixed. However, in this work, we discovered a new behavior of the critical current density; the $J_c$ of a given superconductor is not fixed. We show that the $I_c$ of one superconducting segment can be either increased or decreased by an adjacent segment along the current bias direction, resulting in a modified $J_c$. Additionally, the segments can exhibit equivalent $I_c$ values at low temperature and magnetic field, diverging when either is increased sufficiently. This phenomenon of superconductivity influenced by adjacent superconductors is called the "global critical current effect", and we observe it in polycrystalline superconductors (deposited Nb), single crystalline superconductors ($NbSe_2$), and superconductor/superconductor heterostructures ($NbSe_2/NbSe_2$) demonstrating its universality to some extent. In addition, we find that the critical magnetic field and the critical temperature ($T_c$) of a superconducting segment are independent of adjacent segments, indicating that $I_c$ can be independent of the critical magnetic field. Finally, to demonstrate the global critical current effect in a potential application, we engineered a superconducting diode effect in one segment of a Nb thin film and showed that it can be transferred to another segment where it cannot manifest the effect on its own, offering great potential for integrating complex functionalities within superconducting circuits taking advantage of this non-local behavior. We discuss these observations in the context of known compounding factors such as Joule heating effect, current crowding effect, vortex formation and proximity effect. Overall, these results demand a reexamination of the fundamental mechanisms of superconductors' critical currents, and allow for a new design paradigm of superconducting circuits, electronics, and sensors.

**Experimental Data**

First, superconducting Nb strips with different geometries were fabricated to study their superconducting properties. Fig. 1A shows the optical image of a typical device (device #1), in which the left and right segments have a width of ~0.9 μm, while the middle segment has a width of ~0.45 μm (lower panel of Fig. 1A). The leads connected to the strips have a larger width and they continuously become wider towards the wiring bonding pads. In this structure, there are several different connection configurations possible for characterizing the critical current of each segment. Accordingly, we define two typical measuring configurations: one is the "local configuration", where the segments between the source leads have the same width, and the other is the "global configuration", where the segments between the source leads have differing widths. Taking segment 23 (labelled as $S_{23}$) as an example, when measuring its V-I curve, the voltage leads will always be the leads 2 and 3, but there are multiple choices for the source leads. Sourcing current through the leads 1 and 4 is an example of the "local configuration" whereas sourcing current through the leads 1 and C is the "global configuration".

With these definitions, we first focus on three representative segments in device #1 to study the difference between the two measurement configurations. The V-I curves measured via the 4-probe method are shown in Fig. 1B and 1c. Fig. 1B plots the V-I curves of $S_{23}$, $S_{67}$ and $S_{AB}$ in their local configurations at 3K. These V-I curves exhibit hysteresis with two transitions in each curve within the first quadrant. The larger one represents the critical current ($I_c$), corresponding to the transition from superconducting to normal state in the up-sweep of the applied current; and the smaller one is the return current corresponding to reentering the superconducting state in the down-sweep of the applied current. As expected, the $I_c$ of $S_{23}$ and $S_{AB}$ are similar (~1.4 mA) since they have the same width, whereas the narrow segment $S_{67}$ shows a smaller $I_c$ of 0.4 mA. Fig. 1C plots the *V-I* curves of the same three segments measured at the same time in the global configuration (applying current through leads 1 and C). Surprisingly, all of these three segments now show the exact same $I_c$ with the value of 0.4 mA, which is the same as the $I_c$ of the narrow segment $S_{67}$ measured in the local configuration. In addition, we compared the *V-I* curves measured in the local and global configurations for all three segments, and found that the normal state resistance of each segment remains the same regardless the measurement configuration, even though the $I_c$ can be different for different configurations (see Fig. S1). Moreover, not only these three segments, but all of the segments between leads 1 and C show the same $I_c$ (0.4 mA) in the global configuration, as shown in Fig. S2.

These observations reveal that, in the global configuration the $I_c$ of a superconducting segment can be influenced or determined by other segments. We call this phenomenon the "global critical current effect", which differs from the typical understanding of critical current as a fixed value for a given superconductor. Note that since the $I_c$ can also be influenced by other factors such as sharp corners or turns[20,21] and defects[22,23], the phenomenon we observed highlights that it's necessary to check measurements carefully in superconducting circuits to understand which part dominates the measured critical current. To this end, we propose a methodology based on 2-probe measurement to check the $I_c$ of each segment (detailed in supplementary material and Fig. S3) and determined that the narrow segment always dominates the $I_c$ measured in the global configuration in our results.

We further studied the $I_c$ of the three segments in both configurations as a function of temperature. Fig. 1D plots the temperature dependent $I_c$ of $S_{23}$, $S_{67}$ and $S_{AB}$ in their local configurations. As expected, the $I_c$ of $S_{23}$ and $S_{AB}$ exhibit a similar magnitude and temperature dependence while $S_{67}$ is distinct. However, in the global configuration (Fig. 1E), the $I_c$ of the three segments match perfectly at low temperatures (< 4K), determined by the $I_c$ of $S_{67}$ (blue region). And the $I_c$ of $S_{23}$ and $S_{AB}$ begin to diverge from that of $S_{67}$ at higher temperatures (> 4 K), as indicated by the orange region.

The magnetic field dependent resistance (*R-B* curves) of the different segments in both local and global configurations were also investigated, with Fig. 1F and 1G showing the *R-B* curves for $S_{23}$ and $S_{67}$, respectively. The solid (dashed) lines are measured using local (global) configuration. The *R-B* curves of both $S_{23}$ and $S_{67}$ in both configurations overlap very well at different temperatures. These results indicate that the critical magnetic fields (both $B_{c1}$ and $B_c$) of each segment are independent of the measurement configurations (also see Fig. S4). Moreover, the $T_c$ of $S_{23}$ and $S_{67}$ also do not change when changing measurement configuration (see Fig. S5). These behaviors differ from that of the critical current, which we demonstrated is highly dependent on the measurement configuration. The contradictory phenomena suggests that the zero-field $I_c$ of a superconductor may not be intimately related to its critical magnetic field.

The global critical current effect is not only shown in the polycrystalline deposited Nb film, but also in a single crystalline van der Waals superconductor. As shown in the inset of Fig. 2A, a $NbSe_2$ flake with thickness of ~40 nm was shaped into similar narrow (~0.5 μm) and wide (~1 μm) segments (device #2) as the Nb film (device #1), by Ar milling. Note that the leads on the $NbSe_2$ device are all gold, which is a non-superconducting material, whereas the leads of Nb devices (e.g. device #1) are also Nb. As shown in Fig. S6, the *V-I* curves of the $NbSe_2$ device (device #2) show the same behaviors as in the Nb device (device #1). Specifically, the wide segment ($S_{23}$) in the $NbSe_2$ device shows different $I_c$ in the local vs. global configurations, and its $I_c$ in the global configuration (sourcing leads are 1 and 8 in the case of device #2) is also limited by that of the narrow segment ($S_{67}$). The *V-I* curves of all segments in both local and global configurations at different temperatures were measured, and the corresponding $I_c$ were extracted and plotted as a function of temperature in Fig. 2A. Just as with Nb, here the $I_c$ of the segments ($S_{23}$ and $S_{67}$) in the global configuration are exactly same in the low temperature region, but diverge when temperature is higher than 4.6 K. In addition, we investigated the magnetic field dependence of the critical current for the segments at 4 K in both configurations. As shown in Fig. 2B, at low magnetic fields in the global configuration, $S_{23}$ shows the same $I_c$ as $S_{67}$, which is smaller than the $I_c$ in its local configuration. When increasing magnetic field, the $I_c$ of the segments measured in the global configuration diverge at around 20 mT.

We also investigated the global critical current effect in superconductor/superconductor layered heterostructures; an important platform for studying exotic superconducting phenomena[24–27]. Surprisingly, we observed increased $I_c$ of segments in the $NbSe_2/NbSe_2$ heterostructure (the inset of Fig. 3A, device #5) when the junction was included in the applied current trajectory. The $NbSe_2$ flake with gold leads 1, 2, 3 and 4 is on the top of the other $NbSe_2$ flake with gold leads 5, 6, 7, 8, B and C. Since the width of these flakes are not uniform, we do not define the local configuration, but rather watch the behavior of $I_c$ along a single flake versus across the junction of two flakes (Inset Fig. 3A).

We measured the $I_c$ of $S_{67}$ in different configurations as shown in Fig. 3A. Surprisingly, the $I_c$ of $S_{67}$ increases from 1.82 mA to 2.06 mA when changing from configuration 58 (source current through leads 5 and 8 along the bottom flake) to configuration 18 (source current through leads 1 and 8 across the junction), meaning $J_c$ of $S_{67}$ varies with the measurement configuration. The increased $I_c$ (as well as $J_c$) is also observed for $S_{23}$ when changing from configuration 14 (source current through leads 1 and 4 along the top flake) to configuration 18 (see Fig. S7C). Moreover, all segments in between the leads 1 and 8 ($S_{34}$, $S_{45}$, $S_{56}$ and $S_{67}$) except $S_{23}$ break superconductivity simultaneously, as shown in Fig. 3B and all segments show an increased $I_c$ compared with their respective 2-probe measurements (Fig. S7B).

The temperature dependent (Fig. 3C) and magnetic field dependent (Fig. 3D) $I_c$ of different segments were then measured in different configurations. The $I_c$ (as well as $J_c$) of $S_{67}$ in configuration 18 increases over a wide temperature region compared with $I_c$ (as well as $J_c$) measured in configuration 58. With magnetic field, a helmet-like shape is observed for the $I_c$ (as well as $J_c$) of $S_{67}$ in configuration 18, indicating an enhancement of $I_c$ (as well as $J_c$) at low magnetic fields that disappears above 90 mT. In addition, the simultaneous superconductivity breaking of $S_{45}$ (across the junction) and $S_{67}$ in configuration 18 is observed up to 6.6 K (see the inset of Fig. 3C), and up to 90 mT at 4 K (Fig. 3D). For $S_{67}$ measured in different configurations (both along the bottom flake as well as across the junction), the critical temperature and critical magnetic fields do not change, as shown in Fig. S8. Taken together, these results also show that the $J_c$ of a superconducting segment may not be directly derived from its critical fields.

The observation of the global critical current effect in these different devices suggest that it's possible to transfer critical current related phenomena from one segment of superconductor to adjacent segments where they wouldn't otherwise manifest, like for instance, the superconducting diode effect (SDE). We designed a strip-sawtooth-strip device with Nb (device #3) (see optical images in Fig. 4A), where the sawtooth segment can generate a field induced SDE via its geometry[28,29], and the adjacent two strips are used to receive it. The width of left strip is ~0.9 μm, and the right one is ~1.7 μm. As shown in Fig. 4B, the $I_c$ behave exactly the same as in devices #1 and #2; that is, the $I_c$ of the three representative segments with different geometries exhibit the exact same value in the low temperature region in the global configuration (sourcing current through lead 1 to G in device #3). All other segments between lead 1 and G also have the same $I_c$ in the global configuration measured at 4 K (Fig. S9A). Moreover, it can be clearly seen in the inset of Fig. 4B that, in the global configuration, the $I_c$ of $S_{23}$ and $S_{EF}$ diverge from $S_{67}$ at different temperatures due to the different widths of $S_{23}$ and $S_{EF}$.

We studied the magnetic field dependence of critical currents in device #3. Fig. 4C - 4H plot the colormaps of differential resistance as a function of magnetic field and applied current at 4 K. Fig. 4C-4E are the results of $S_{23}$, $S_{67}$ and $S_{EF}$ measured in their corresponding local configurations, whereas Fig. 4F - 4H are those measured in the global configuration. As can be seen from the colormaps, both the magnitude and tendencies of critical currents of $S_{23}$ and $S_{EF}$ vary significantly with measurement configuration, whereas the results of $S_{67}$ remain the same. The $I_c$ were then extracted and plotted as a function of magnetic field in Fig. S10, where the $I_c$ of all three segments in the global configuration has same magnitude and tendency in the low magnetic field region. This proves that the sawtooth segment dominates the critical current behavior in the global configuration at 4K, as designed. In addition, as magnetic field increases, the $I_c$ of $S_{EF}$ diverge from that of $S_{67}$ above 37 mT, whereas $S_{23}$ doesn't exhibit this behavior within 80 mT (Fig. S10A). Similar with the temperature induced divergence, the critical current of $S_{EF}$ with larger width diverges easier from that of $S_{67}$ than that of $S_{23}$.

Finally, to compare the SDE in these segments in different measurement configurations, the diode efficiencies ($\eta = \frac{|I_{c+}-|I_{c-}||}{I_{c+}+|I_{c-}|}$) of the three segments were calculated and plotted in Fig.4I, 4J and Fig. S10. In the global configuration, as shown in Fig. 4I, the $\eta$ of the three segments overlap perfectly at low magnetic fields, showing the same diode behavior with a maximum of 14% at ~17 mT. As magnetic field increases, the $\eta$ of $S_{EF}$ diverges from that of $S_{23}$ and $S_{67}$, dropping quickly to near 0. To investigate the transfer of the SDE, the comparisons of $\eta$ with different configurations were also plotted in Fig. 4J ($S_{EF}$) and Fig. S10E and S10F ($S_{23}$ and $S_{67}$). It can be clearly seen that $\eta$ of $S_{23}$ and $S_{EF}$ changed from near zero to 14% when changing from a local to a global configuration, demonstrating that the idea of transferring SDE (from the sawtooth segment in this case) to adjacent segments (a strip segment in this case) where it can't otherwise manifest, is indeed possible. We speculate that, through the global critical current effect, other critical current related phenomena may also be able to be transferred to adjacent parts. As such, the design of superconducting devices can be reconsidered; by separating the functionally generating parts and the detecting parts, it is possible to integrate multiple phenomena into a single device and realize complex superconducting applications.

**Discussion**

We briefly summarize several key phenomena observed in both deposited polycrystalline and single-crystalline superconductors as well as in superconductor/superconductor heterostructures, where segments of a superconductor can show modified $I_c$ with measurement configuration: 1) A single segment can have different $I_c$, as well as $J_c$, values depending on the measurement circuit, 2.) but its critical magnetic field and $T_c$ remain unchanged. 3) Different segments can have the same $I_c$ value at low temperatures and magnetic fields, diverging as either increase and 4) the superconducting diode effect can be transferred from one segment to another segment, even if the other segment couldn't manifest an SDE on its own.

The unchanged critical magnetic field and $T_c$ in different measurement configurations of a superconducting segment indicate that the superconducting gap of each segment is not influenced by the measurement configuration. Only $I_c$, as well as $J_c$, being modified by other segments, implies that the intimate link between $J_c$ and the critical magnetic field[2,3] may need to be reexamined, and the mechanism of the current driven superconducting breaking may need to be revisited. Below we discuss several factors related to critical current and the breaking of superconductivity.

First, the Joule heating effect is widely used in the explanation of critical current related phenomena such as the hysteresis of V-I curves of superconductors[30–32]. Conventionally, when a superconducting segment with lower $I_c$ breaks superconductivity, due to the Joule heating effect, one expects a significant amount of heat to be locally generated that dissipates to the adjacent segments, raising their local temperatures and thereby breaking superconductivity across the entire superconductor simultaneously. This effect also means that the $I_c$ of adjacent segments will only decrease, due to their increased local temperature. However, we find that the simultaneous superconductivity breaking of different segments does not happen in several cases. For example, we measured the critical current of different segments ($S_{B5}$, $S_{56}$, $S_{67}$) on the bottom NbSe$_2$ flake in configuration C8 (source current through leads C and 8) in the NbSe$_2$/NbSe$_2$ heterostructure (device #5), as shown in Fig. S7A. Contrary to the expectation from local heating, we find that the different segments do not break superconductivity simultaneously (Fig. S7A). We also observe this same behavior in a pristine superconducting NbSe$_2$ flake (Fig. S11) as well as a kagome superconductor KV$_3$Sb$_5$ flake (Fig. S12).

Moreover, instead of a decreased $I_c$, we found an increased $I_c$ of segments in the NbSe$_2$/NbSe$_2$ heterostructure and the KV$_3$Sb$_5$ flake, when other segments with lower $I_c$ are included in the applied current trajectory. For example, as shown in Fig. 3A, the $I_c$ of $S_{67}$ in the NbSe$_2$/NbSe$_2$ heterostructure is increased when current is applied across the junction (configuration 18) compared to when current is applied only along the bottom flake (configuration 58). The increased $I_c$ is also found for $S_{23}$ when current is applied across the junction vs along only the top flake (see Fig. S7C). In the KV$_3$Sb$_5$ flake (device #6), the $I_c$ of $S_{DE}$ is increased when including narrower segments along the current trajectory. Taken together, these results imply that the Joule heating effect is not the driving factor of the global critical current effect.

Another important factor, the current crowding effect, is considered in superconductors with sharp corners or turns, which can lead to a reduced critical current[20,21,33]. This effect arises from a non-uniform current density distribution which increases around sharp corners or turns, making it easier for the nucleation of vortices and antivortices, resulting in a reduced $I_c$ near those sharp corners or turns[33,34]. However, as discussed above, the $I_c$ in some of our cases can increase rather than decrease, contrary to this expectation. At this time, the current crowding effect does not consider the influence of one superconducting segment on adjacent segments, as we have shown is important in the global critical current effect. Hence, the observed phenomena in this work cannot be explained by the current crowding effect.

Also, the formation and motion of flux vortices is an important mechanism for current and magnetic field induced superconductivity breaking[1,35]. It has been shown that the critical current of a superconductor is determined by its critical magnetic field and microstructure, where the former governs the formation of flux vortices and the latter determines the pinning force of those vortices. Once the Lorentz force on the vortices exceeds the pinning force, the vortices start moving, thereby breaking superconductivity[1,35]. It's been previously reported that the $J_c$ (as well as $I_c$) of a superconductor can be increased due to, for instance, increased pinning centers arising from local modulation of the microstructure in the superconductor[36,37]. However, in our experiments, the microstructure of the measured superconducting segments is not modulated when changing measurement configuration. In addition, as demonstrated above, the $I_c$ of a segment varies with measurement configuration, but the critical field does not, which is contrary to the expectation from flux vortex formation and motion. These observations indicate that the conventional flux vortices formation and motion theory is not adequate to explain the global critical current effect.

Finally the proximity effect is can possibly lead to an increase $I_c$ (as well as $J_c$) of a superconductor when it is connected to another one with a larger $I_c$[38,39]. However, in the NbSe$_2$/NbSe$_2$ heterostructure (device #5) for example the $I_c$ of $S_{67}$ measured in configuration 18 (current applied across the junction) is larger than the $I_c$ of any segment in between the leads 1 and 8 measured in their respective 2-probe measurement, as indicated by the 2 mA reference line in Fig. 3A, 3B and Fig. S7. Moreover, the $I_c$ of $S_{BC}$ and $S_{DE}$ in the KV$_3$Sb$_5$ flake shows increased values when narrower segments are included into the applied current trajectory (see detail discussion in supplementary material). Hence the proximity effect cannot explain the observations either.

Overall, the exact fundamental mechanism of the global critical current effect is still unclear. It may involve a combination of multiple factors including those addressed above, as well as potential new theories that consider the interaction between different superconducting segments. In addition, it's also possible that due to the competition of the factors, the dominant factor accounting for the global critical current effect might be different in different superconductors or superconducting structures. Further theoretical and experimental studies are needed to understand the fundamental mechanism of the global critical current effect, including measuring more superconducting materials in a variety of geometries and limits.

In conclusion, we report the discovery of the global critical current effect of superconductivity, in very different materials, demonstrating its universality to some extent. It shows that the critical current of a superconductor is not only dependent on its own properties, but can also be influenced by its adjacent superconductors along the applied current trajectory. Specifically, superconductivity of superconductors with different geometries can be broken simultaneously by the applied current, which can be disrupted at high temperature or magnetic field. On the contrary, the critical magnetic fields and critical temperatures of the superconductors remain unaffected, meaning the critical current is not as tied to the critical magnetic field as previously thought. The fundamental mechanism of the current induced superconductivity breaking is still veiled at this stage. Nevertheless, regardless of the mechanism, this effect highlights that more attention should be paid in superconductivity measurement and analysis, as a superconducting segment can exhibit critical current related properties that deviate significantly from its intrinsic state. Moreover, using the global critical current effect, we demonstrated that superconductors can transfer their critical current related properties to each other; we transferred the superconducting diode effect to a segment where it can't manifest otherwise. On one hand, this phenomenon calls for careful scrutiny when investigating the mechanism of an observed SDE, as it may originate from segments other than the measured one; on the other hand, it implies that other critical current related phenomena may also be able to be transferred to other segments, providing new avenues for designing superconducting circuits and electronics exploiting this non-locality.

**Acknowledgments:**

The authors acknowledge Kun Jiang for valuable discussion, and Houssam el Mrabet Haje for commenting on the manuscript.

**Funding:** H.W. acknowledges that this research was supported by from NWO Talent Programme VENI financed by the Dutch Research Council (NWO) VI.Veni.222.380. Y.W. acknowledges the support from NWO Talent Programme VENI financed by the NWO, project no. VI.Veni.212.146. M.N.A acknowledges support from the NWO Talent Programme VIDI financed by the NWO VI.Vidi.223.089, the Kavli Institute Innovation Award 2023, the Kavli Institute of Nanoscience Delft, and the research program "Materials for the Quantum Age" (QuMat, registration number 024.005.006) which is part of the Gravitation program financed by the Dutch Ministry of Education, Culture and Science (OCW).

**Author contributions:** H.W. and Y.W. conceived and designed the study. H.W. and Y.W. fabricated the devices and performed all the measurements. H.W. carried out the data analysis. M.N.A is the Principal Investigator. All authors contributed to the preparation of manuscript.

**Competing interests:** The authors declare that they have no competing interests.

**Data and materials availability:** The data that support the findings of this study are available from the corresponding author upon reasonable request.

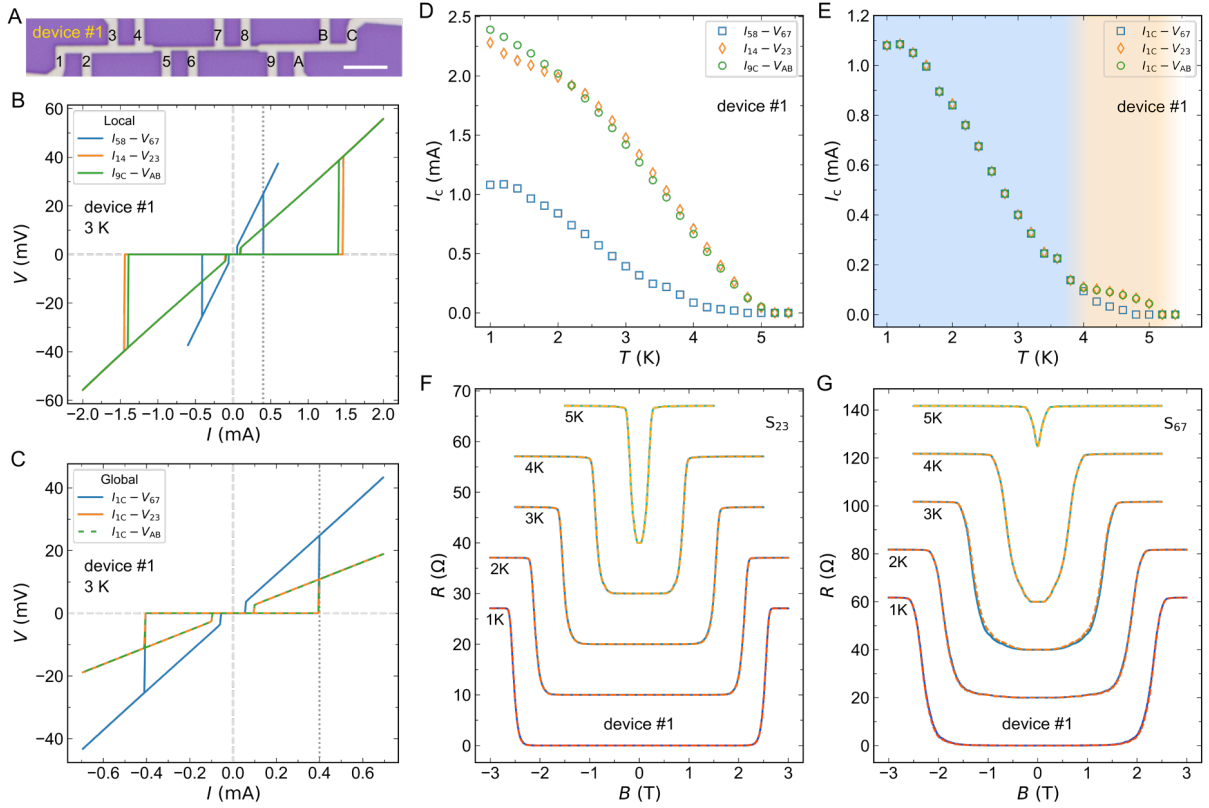

**Fig. 1. The global critical current effect in the Nb thin film (device #1). (A)** The optical images of the Nb strip (device #1). The strip contains three segments with different widths, where the left and right ones are ~0.9 μm and the middle one is ~0.45 μm. The scale bar is 5 μm. **(B)** The *V-I* curves of $S_{23}$, $S_{67}$ and $S_{AB}$ measured in their respective local configurations at 3 K, which shows that the $I_c$ of $S_{23}$ and $S_{AB}$ are similar, and are larger than that of $S_{67}$. **(C)** The *V-I* curves of $S_{23}$, $S_{67}$ and $S_{AB}$ in the global configuration at 3 K (source leads are 1 and C), where the $I_c$ of these segments match perfectly. **(D)** Temperature dependent $I_c$ of $S_{23}$, $S_{67}$ and $S_{AB}$ measured in their respective local configurations. **(E)** Temperature dependent $I_c$ of $S_{23}$, $S_{67}$ and $S_{AB}$ in the global configuration, where they match perfectly at low temperatures and start to diverge at 4 K. **(F)** Temperature dependent *R-B* curves of $S_{23}$ measured in the local (solid lines) and global (dashed lines) configurations (applying an a.c. current of 1 μA). The curves of different configurations overlap very well, indicating the critical magnetic fields (both $B_c$ and $B_{c1}$) are independent with the measurement configurations, unlike the critical currents. The magnetic field direction is out-of-plane. **(G)** Temperature dependent *R-B* curves of segment $S_{67}$ measured in the local (solid lines) and global (dashed lines) configurations (applying an a.c. current of 1 μA). Same with the result of $S_{23}$, the critical magnetic fields of $S_{67}$ are independent with the measurements configurations. The magnetic field direction is out-of-plane.

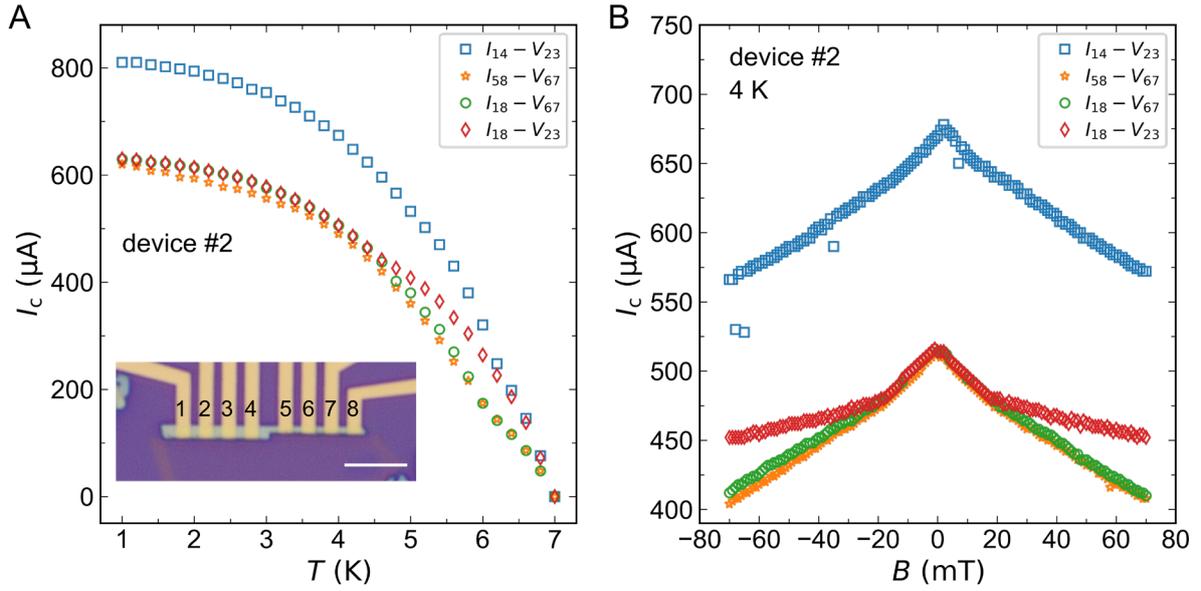

**Fig. 2. The global critical current effect in a single crystal superconductor NbSe$_2$ (device #2) and its magnetic field dependency.** **(A)** Temperature dependent $I_c$ of S$_{23}$ and S$_{67}$ measured in the local and global configurations. The $I_c$ of S$_{23}$ and S$_{67}$ in the global configuration match very well at low temperatures and start to diverge at 4.6 K. The inset is the optical image of the NbSe$_2$ strip (device #2). The contacts are Ti/Au, and the width of wide segment is ~1 μm and the width of the narrow segment is ~0.5 μm. The scale bar is 5 μm. **(B)** Magnetic field dependent $I_c$ of S$_{23}$ and S$_{67}$ measured in the local and global configurations. Similar to the temperature dependent results, the $I_c$ of S$_{23}$ and S$_{67}$ in the global configuration match perfectly at low magnetic fields, whereas they start to diverge at around 20 mT. The magnetic field direction is out-of-plane.

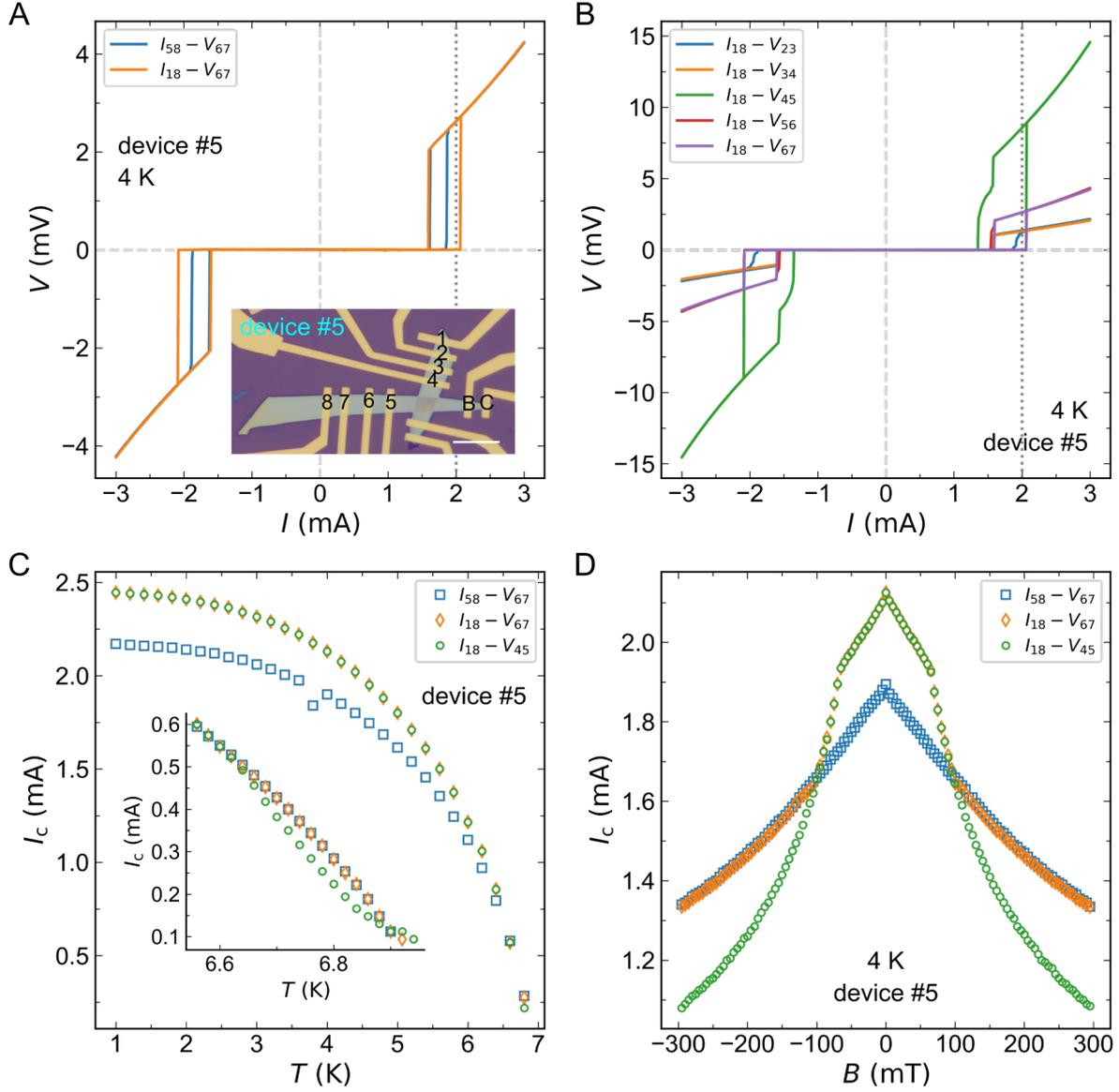

**Fig. 3. The global critical current effect in a NbSe$_2$/NbSe$_2$ heterostructure (device #5).** **(A)** The *V-I* curves of S$_{67}$ measured in configuration 58 (source leads are 5 and 8) and configuration 18 (source leads are 1 and 8). The inset is the optical image of the NbSe$_2$/NbSe$_2$ heterostructure (device #5). The scale bar is 10 μm. **(B)** The *V-I* curves of all the segments in configuration 18, where all of them except S$_{23}$ breaks superconductivity simultaneously. **(C)** Temperature dependent $I_c$ of S$_{45}$ (across the junction) and S$_{67}$ in configuration 18, and $I_c$ of S$_{67}$ in configuration 58. The $I_c$ of S$_{45}$ and S$_{67}$ in configuration 18 match perfectly at low temperature, and they start to diverge at 6.6 K, revealed by the detailed measurement shown in the inset. **(D)** Magnetic field dependent $I_c$ of S$_{45}$ and S$_{67}$ in configuration 18, and $I_c$ of S$_{67}$ in configuration 58 at 4K. Similar to the temperature dependent results, $I_c$ of S$_{45}$ and S$_{67}$ in configuration 18 match very well at low fields and start to diverge at 90mT. The magnetic field direction is out-of-plane.

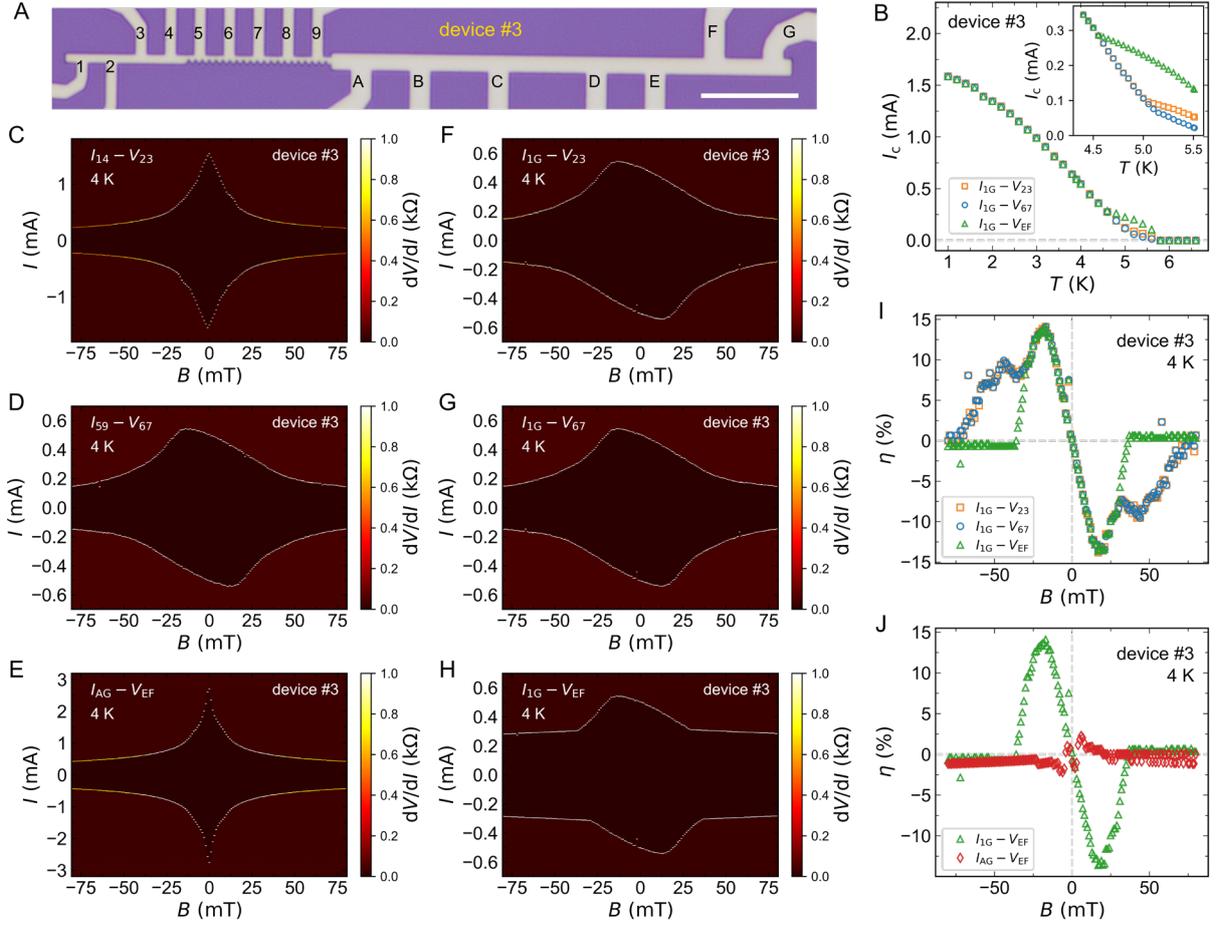

**Fig. 4. The transfer of superconducting diode effect (device #3). (A)** The optical image of the Nb strip-sawtooth-strip device (device #3), where the left strip has a width of ~0.9 μm and the right strip has a width of ~1.7 μm. The scale bar is 10 μm. **(B)** Temperature dependent $I_c$ of $S_{23}$, $S_{67}$ and $S_{EF}$ measured in the global configuration (source leads are 1 and G). The inset is the detailed measurement where the divergence of $I_c$ begins. **(C-H)**. Color maps of differential resistances as a function of out-of-plane external magnetic fields and applied current, the results are obtained from $S_{23}$, $S_{67}$ and $S_{EF}$ in both local **(C-E)** and global **(F-H)** configurations at 4 K. **(I)** The $\eta$ of $S_{23}$, $S_{67}$ and $S_{EF}$ obtained from **(F-H)**, which match very well at low magnetic fields. The $\eta$ of $S_{EF}$ starts to diverge from others around 30 mT, and falls to 0 quickly. **(J)** The $\eta$ of $S_{EF}$ obtained from **(E)** and **(H)**. It only shows a small $\eta$ value in the local configuration, whereas a much larger one in the global configuration (with the sawtooth segment in the same circuit), demonstrating the transfer of superconducting diode effect.

14